\documentclass{emulateapj}
\usepackage{epsfig}

\def\arcmin{$^{\prime}$}

\shorttitle{Subgroups of Globular Clusters}
\shortauthors{Woodley \& Harris}

\begin{document}

\title{Possible Subgroups of Globular Clusters and Planetary Nebulae in NGC 5128}

\author{Kristin A.~Woodley}
\affil{Department of Physics \& Astronomy,
  University of British Columbia, Vancouver BC V6T 1Z1, Canada} \email{kwoodley@phas.ubc.ca}

\author{William E.~Harris}
\affil{Department of Physics \& Astronomy, McMaster University,
  Hamilton ON  L8S 4M1, Canada}
\email{harris@physics.mcmaster.ca}

\begin{abstract}
We use recently compiled position and velocity data for the globular cluster and planetary nebula subsystems in NGC 5128, the nearby giant elliptical, to search for evidence of past dwarf-satellite accretion events.  Beyond a 10\arcmin\ ($\sim11$ kpc) radius in galactocentric distance, we find tentative evidence for 4 subgroups of globular clusters and 4 subgroups of planetary nebulae.  These each have $> 4$ members within a search radius of 2\arcmin\ and internal velocity dispersion of $\lesssim40$ km s$^{-1}$, typical parameters for a dwarf galaxy.  In addition, 2 of the globular cluster groupings overlap with 2 of the planetary nebulae groupings, and 2 subgroupings also appear to overlap with previously known arc and shell features in the halo light.  Simulation tests of our procedure indicate that the probability of finding false groups due to chance is $<1\%$.
\end{abstract}


\section{Introduction}
\label{sec:intro}
In the last decade or so, the presence of stellar streams within nearby galaxies has provided direct evidence for the continual build-up of massive galaxies from the accretion of small neighbouring satellites.  For example, in the Milky Way, there is the Sagittarius stream \citep{ibata97,ibata01a,majewski03,martinez04}, and in M31, a number of new streams have been recently discovered within the {\it Pan-Andromeda Archelogical Survey} \citep{mcconnachie09}, adding to the list of already known streams in that galaxy \citep{ibata01b}. As well, striking surface brightness features tracing a galaxy's merger history have been found in NGC 5907 \citep{shang98}, NGC 891 \citep{mouhcine10a}, NGC 4093 \citep{martinez08}, NGC 253 and NGC 5236 \citep{malin97}, and UGC 10214 \citep{forbes03}, among others.  Past work has shown that it is possible to use streams to trace the orbits of satellite galaxies that had merged with the Milky Way in the distant past \citep{lyndenbell95,yoon02}.  \cite{lyndenbell95} suggest that the tidal debris stripped out of the parent dwarf galaxy  will not undergo much dynamical friction because its mass content will be low. 

In systems beyond our Local Group galaxies, finding these traces of old satellites becomes more difficult.  But in addition to the field-star integrated light, it is possible to use other old objects such as globular clusters (GCs) and planetary nebulae (PNe) to provide evidence for the presence of accretion remnants \citep{forte82,muzzio87,cote98,cote02,hilker99}.  As well, \cite{pipino07} have modelled the GC assembly history in massive galaxies, and showed that in addition to the GCs that formed along with the galaxy halo, a significant fraction of GCs that make up the massive galaxies were likely accreted from dwarf galaxies in later times.  
Indeed, subgroups of GCs have been found within the halos of massive galaxies or have been associated with the stellar streams in the Sagittarius stream \citep{ibata94,ibata97,bellazzini03} and in M31 \citep{perrett03,mackey10}.

For our search, we target NGC 5128 (Centaurus A) which is a giant elliptical galaxy at a distance of $3.8\pm0.1$ Mpc \citep{harris10}.  While this galaxy is quite close, it is still beyond the distance where detecting faint stellar streams is an easy task.  Previous work by \cite{malin83} and \cite{peng02} has shown there are existing complex shell structures that surrounds the central regions out to 15 kpc, attributed to the many accretions of small neighbouring galaxies.  There are also HI shells \citep{schiminovich94} and a prominent warped dust lane \citep{graham79} is also present in the central region where gas and dust is settling into the central potential well.  There is evidence for young stars \citep{rejkuba01,rejkuba02,ellis09} that may be aligned with the radio jet.  This star formation could have been triggered by the jet colliding with cloud material brought into NGC 5128 via a merging event.  

NGC 5128 has 607 confirmed GCs, via radial velocity measurements \citep{vandenbergh81,hesser84,hesser86,harris92,peng04b,woodley05,rejkuba07,beasley08,woodley10a,woodley10b} and/or resolved images from the {\it Advanced Camera for Surveys} on the {\it Hubble Space Telescope} \citep{harris06,mouhcine10b}.  Of these, 563 have measured radial velocities.  Their mean measurement uncertainty is 42 km s$^{-1}$, though the range of uncertainties is quite large, with $96\%$ of the population having uncertainties $< 200$ km s$^{-1}$.   
The currently knowns GCs are distributed in galactocentric radius out to 50\arcmin\ (where 1\arcmin\ $\sim1.1$ kpc at the distance of NGC 5128), although the inner 5\arcmin\ is sparsely populated because of the obscuration of the well known central dustlane.  There are also 780 confirmed PNe with positional and radial velocity data \citep{peng04a}, which can be used as stellar tracers in the same way.
 
\section{Group Identification Technique}
\label{sec:id}

\subsection{Globular Cluster Observational Dataset}
\label{sec:obsdata}

From both the GC and PNe databases, we attempt to identify genuine subgroups in the classic way by looking for objects that are close to each other in both position and velocity.  Our basic search approach follows that of \cite{perrett03} for the GCs in M31.  We look first at the GCs.  Here the NGC 5128 GC data face us with the additional difficulty beyond the situation in M31: not only are the positions and velocities seen only in projection, but the velocity measurement uncertainties (averaging more than 40 km s$^{-1}$) are usually larger than the expected $\sim 20$ km s$^{-1}$ true, physical velocity dispersion of a typical satellite dwarf in the Local Group \citep{geha10}.  For this reason, we view our present work as only a preliminary step towards finding guenuine physical subgroups.  For the same reason, we suggest that at this stage, only rather simple group-finding search procedures are justified.  Logical next steps would include higher-precision velocity data, and imaging to search for faint surface brightness features.

GC metallicity or color might in principle be additional search criteria, as it may be expected that GCs within one accreted system may share similar enrichment processes, but these properties are less useful in practice.  In our sample of GCs, the metallicities are either from a color transformation or simply based on a color division which we take from \cite{woodley07,woodley10a,woodley10b}.
We have classified the GCs as being either metal-poor or blue ($[Fe/H] < -1.0$) or as metal-rich or red ($[Fe/H] \geq -1.0$) following previous work on the GC system of NGC 5128 \citep{harris04b,woodley05,woodley07} or as metal-rich or red if ($B-I$)$\geq 2.072$ and metal-poor or blue if ($B-I$)$< 2.072$, following \cite{peng04b}. 
The division between red and blue GCs is thus not homogeneously determined and depends on a variety of methods as well as a number of different photometric studies.  Of these GCs, 291 are classified as metal-poor, 292 as metal-rich, and 24 GCs have insufficient photometry for any of the above transformations.
In addition, while there is evidence that the metallicities, on average, of GCs in dwarf galaxies are more metal-poor than in more massive galaxies \citep{sharina10}, 
there is evidence that a range of metallicities of GCs can exist within a single dwarf galaxy, for example, in the dwarf galaxies IC10 and UGCA86 \citep{sharina10}.  As well, there can exist a spread in the metallicities of the dwarf galaxies themselves \citep[see the review of][]{mateo98}.  This information, taken together, makes it quite difficult to incorporate a defined range of acceptable color or metallicity within a subgroup as a search parameter (we note that \cite{perrett03} also ignored metallicity as a search parameter).

We searched for GC subgroups in the outer regions of NGC 5128, beyond 10\arcmin. Inside this radius, there would be a high fraction of false subgroups because the density of objects is high, and the biggest source of finding false subgroups is due to the accidental 2-dimensional projection of objects that are actually far apart along the light of sight.  Additionally, it is the outer halo that should still preserve coherent traces of accreted satellites because the much larger dynamical times there prevent them from being fully phase-mixed \citep{bullock05}.

As mentioned above, we use essentially a nearest-neighbour approach with two linking criteria.  For each GC outside a 10\arcmin\ galactocentric distance, we calculate the separation in projected linear distance between it and all other GCs in the catalog (the boundary can therefore extend slightly inwards of 10\arcmin) and group the GCs if these are within our imposed linking length.  We also included the GC velocity and velocity uncertainties in our search such that two GCs would match if the range of its velocity plus or minus its velocity uncertainty can be subtracted from another objects velocity such than their absolute difference is less than 20 km s$^{-1}$.  For example, objects with velocities $500\pm30$ km s$^{-1}$ and $460\pm20$ km s$^{-1}$ would be linked.  To help avoid false group detections based on poorly measured velocities, we only considered GCs whose velocity uncertainties were less than an imposed limit.  We also combined groups that were found to have common members, which allows for groups that may exist in a chain-like structure.

A search using a spatial linking length of 1\arcmin\ did not yield any subgroups with more than 4 members.  When we increased the linking length to 2\arcmin, which is a reasonable size for a subgroup from a dwarf galaxy, we did indeed find plausible subgroups. 
With a linking length of 2\arcmin, as well as allowing for velocity uncertainties of up to 40 km s$^{-1}$, we have detected a plausible subgroup of GCs including 6 group members.  When we include GCs in our search that have measured velocity uncertainties up to 50 km s$^{-1}$, we increase the number of group members by 1, and call this a possible group of 7 members, labelled GC 1.   When we increase the allowed velocity uncertainty to 80 km s$^{-1}$, we end up finding 3 additional subgroups, each consisting of 5 members, which we label groups GC 2, GC 3, and GC 4.  Many of these are likely to be included simply because of the large velocity allowance and we cannot be sure that these are all real groups.  If we set our velocity uncertainty limit to 40 km s$^{-1}$, but increase our linking length to 3\arcmin, we increase the number of members in our subgroup from 6 to 10 (included as possible members of group GC 1).  Many of these included GCs may not be real members. The possible subgroups are plotted in Figure~\ref{fig:subgroups_all} and are listed in Table~\ref{tab:subgroups_gc} with columns of the GC ID, group ID, $R.A.$ and $Decl.$ in J2000 coordinates, the $[Fe/H]$ value, their galactocentric radius, $R_{gc}$, and their measured radial velocity, $v_r$, and associated uncertainty, $\sigma_{v_r}$.  The radial velocities and uncertainties are weighted values determined from all previous measurements in the literature.  These values are obtained from \cite{woodley07}, \cite{woodley10a} and \cite{woodley10b}.

The image of NGC 5128 in Figure~\ref{fig:subgroups_all} was obtained with the {\it Mosaic II} optical CCD camera on the 4 meter {\it Blanco} telescope at the {\it Cerro Tololo Inter-American Observatory}.  It was provided to us in its reduced form graciously by Eric Peng with its reduction described in \cite{peng02}.  We have smoothed the image using a $300\times300$ pixel box and subtracted it from the original image.  To enhance the visibility of the low-surface brightness shells and arc structures and preserve some of the dominant bulge light, we added back $15\%$ of the smoothed image.  We find that in a general sense, the candidate subgroups that we have identified fall in the same radial zones of galactocentric distance where the shells are found.  However, a stronger indication of possible connections is seen along the isophotal major axis of the galaxy ($35^{\it o}$ and $215^{\it o}$ E of N, \cite{dufour79}).  Toward the southeast (lower right in Figure~\ref{fig:subgroups_all}), group GC 1 falls on the irregular plume of light, while onthe opposite side to the northwest, group GC 3 falls on or close to the two of the visible arcs.  The remaining two groups, GC 2 and GC 4, do not seem to be projected on any known surface brightness features.

\subsection{Planetary Nebula Observational Dataset}
\label{sec:obsdata_pne}

\begin{figure}
\plotone{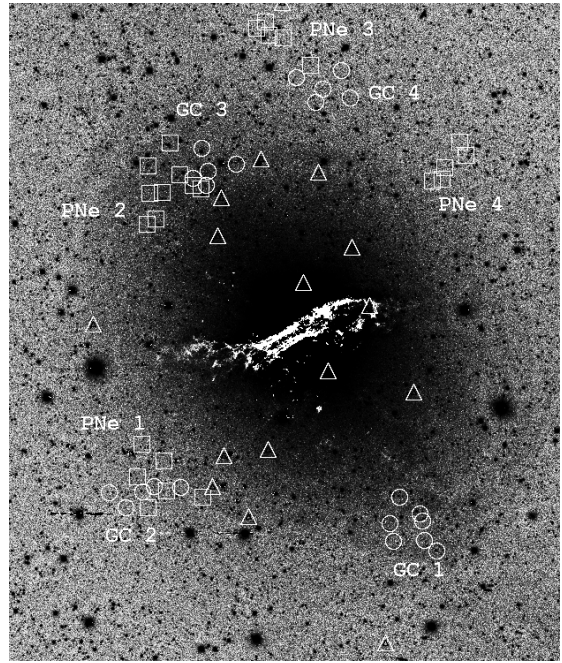}
\caption{We show the positions of the potential groups of GCs and planetary nebalae in NGC 5128.  The group labelled GC 1 ({\it circles}) are the 7 members found with a linking length of 2\arcmin\ with allowed velocity uncertainties of 50 km s$^{-1}$.  We also plot the 3 additional GC subgroups that are found when GCs with velocity uncertainties of up to 80 km s$^{-1}$ are included, labelled GC 2, GC 3 and GC 4 ({\it circles}). 
The 4 plausible PNe subgroups are also shown, groups PNe 1, PNe 2, PNe 3 (note one PNe is off the plot), and PNe 4 ({\it squares}).  The locations of the friendless PNe determined in Section~\ref{sec:orbits} with $N=10$ and $\sigma_N=3$ (see text for a detailed description of these parameters) are also plotted ({\it triangles}).
The image of NGC 5128 is $\sim25$\arcmin\ across with North up and East to the left, and objects in the outer regions are not shown (see text).} 
\label{fig:subgroups_all}
\end{figure}

We have performed the same subgroup finding algorithm on the available PNe data in the literature.  There are 780 PNe with radial velocity measurements \citep{peng04a, hui95} in NGC 5128, which have been detected by the [OIII] emission line.  While there are no velocity uncertainties presented in the literature for the PNe, we have assumed a radial velocity uncertainty of 20 km s$^{-1}$ for each object, which is the typical $rms$ velocity error presented in \cite{peng04a}.  Using our same subgroup finding criteria as for the GCs (linking length of 2\arcmin, velocity difference of 20 km s$^{-1}$ plus the inclusion of the velocity uncertainties, at a minimum distance of 10 \arcmin\ from the center of the galaxy for the central group member), we have found 4 potential groups.  These groups are plotted in Figure~\ref{fig:subgroups_all} and listed in Table~\ref{tab:subgroups_pne} which provides their ID from \cite{peng04a}, their group ID, $R.A.$ and $Decl.$ in J2000 coordinates, the galactocentric radius, $R_{gc}$, their radial velocity measurement, $v_r$,  and assumed 20 km s$^{-1}$ velocity uncertainty.   Out of the 4 subgroups of PNe discovered, 3 have group sizes of $5-6$, while the remaining group has 9 members. These groups have been labelled PNe 1, PNe 2, PNe 3, and PNe 4.  

\subsection{Properties of Subgroups}
\label{sec:properties}

The mean velocity and $rms$ velocity dispersions for each group are listed in Table~\ref{tab:properties}.   
The velocity dispersion for each subgroup ranges from $29-43$ km s$^{-1}$ for the GC subgroups and from $12-24$ km s$^{-1}$ for the PNe subgroups, within a reasonable range for a plausible group.

Our results indicate there are 3 possible overlaps between the GC and the PN subgroups.  These are PNe 1 with GC 2 (Group 1), PNe 2 with GC 3 (Group 2), and PNe 3 with GC 4 (Group 3).  Examining the average velocities for each of these subgroups, the average radial velocity differences are $\Delta(<v_{GC}>-<v_{PNe}>) = 39$ km s$^{-1}$ for Group 1, $-136$ km s$^{-1}$ for Group 2, and $0$ km s$^{-1}$ for Group 3.  The difference in average velocities for Group 1 and Group 3 are within the expected uncertainty range of the measured velocities, which is $42$ km s$^{-1}$ for the average GC.  The subgroups of Group 2, however, do not appear to have strong velocity connections, and are likely either not real groups or are overlapping by chance in projection.  In Sections~\ref{sec:nonparam} \&~\ref{sec:overlap}, we analyze the probability that the overlapping can occur by chance.

In very rough terms we can estimate the total luminosity of the underlying satellite galaxy that would consist of either $\sim 6$ GCs or 6 PNe, typical of what we have found in our subgroups.  
Considering first the GC subgroups, we {\it assume} a specific frequency, $S_N$, which is the number of GCs per unit $M_V = -15$ galaxy luminosity \citep{harris81}.  We adopt $S_N=3$, typical for a normal faint galaxy, though note that \cite{peng08} show that the range of $S_N$ for dwarf galaxies spans 1-20. With these inputs, we obtain $M_V = -15.8$ mag and a surface brightness of $24$ mag per arcsec$^{2}$ assuming the galaxy is spherically shaped with a 2\arcmin\ radial extent.
For a PN subgroup of $\sim6$ members, we can estimate the bolometric luminosity of the underlying satellite galaxy with the luminosity specific PN number, $\alpha = N_{PN}/L_{tot}$, where $N_{PN}$ is the total number of PNe and $L_{tot}$ is the total luminosity \citep{jacoby80}.  \cite{ciardullo05} shows $\alpha_{0.5} \sim 2 \times 10^{-9}$ for galaxies in the range of M$_B \sim -14 $ to $-16$ (please see their Figure~1).  However, in the case of NGC 5128, the PNe data extend at least 1.5 mag down the luminosity function \citep{peng04a}, and we thus estimate $\alpha_{1.5} \sim 11.5 \times 10^{-9}$, to obtain $L_{tot} = 5.2 \times 10^8 L_\odot$.  We obtain $M_V = -16.2$ mag after a bolometric correction of -0.85 \citep{buzzoni06}.   Assuming the galaxy is spherical with a radial extent of 2\arcmin, we obtain a surface brightness of $24$ mag~per arcsec$^{2}$. Both estimates of the surface brightness of a potential underlying dwarf galaxy are very rough guides due to the input assumptions, and we do not think this test is capable of ruling out whether or not we should see these structures in previously obtained deep images of NGC 5128.  We appear to be finding mild traces of substructure that are not hugely above the statistical noise. 

\subsection{Scrambling the Velocities}
\label{sec:nonparam}

Our first step towards assessing how real our candidate groups in Table~\ref{tab:subgroups_gc} are is through a non-parametric test.  By taking our GC positions in 2-dimensional space, we randomly assign each GC a unique and real velocity and velocity uncertainty from the GC catalog.  We then perform our search for subgroups with a linking length of 2\arcmin, a velocity difference of 20 km s$^{-1}$ plus the inclusion of velocity uncertainties up to $40$ km s$^{-1}$, at a minimum distance of 10 \arcmin\ from the center of the galaxy for the central group member.  Out of 1000 trials, we found that 23.2\% of cases found one group containing at least 4 members.  In 1.8\% of the cases, the trials found 2 groups with at least 4 members, and in 1 case (0.1\%), there were 3 groups found.   This indicates that it is likely that one of the GC candidate subgroups may be a false group.  We did not find a case, however, in which 4 candidate groups arose with this methodology.

We examined closely the groups in this test that had more than one group per trial.  In the 18 cases that found 2 groups, there were 2 trials in which the groups overlapped in projected space.  In the trial consisting of 3 groups, 2 of these groups overlapped.  This indicates, that while finding more than 1 group in a trial was rare, when it did occur, there was an approximate 15\% chance they would overlap.  Since these groups were all from the GC population, none of the subgroups that overlapped had similar average radial velocities.

Finally, we randomly assigned each GC a unique and real velocity and velocity uncertainty from the GC catalog, but this time, we incorporated GCs with velocity uncertainties up to $80$ km s$^{-1}$.  Our results indicate that out of 1000 trials,  at least 4 false groups were found in 1.5\% of the cases.  This simple test indicates that it is not likely that {\it all} 4 GC subgroups are false, but it is quite probable that some may indeed by false.  

\section{Simulating Globular Cluster Systems}
\label{sec:sim}

\subsection{Searching for False Subgroups}
\label{sec:false}

Here, we assess how often our search procedure would turn up false groups of similar apparent quality from purely random distributions.  It is similarly important to examine how often our procedure would successfully find a real subgroup within the data.

To carry out tests of the procedure, we generated simulated GC systems in which the particles are located randomly within a spherically symmetric space distribution following a 3-dimensional density power law $\rho \sim r^{-n}$, with $n(3D)=2.5$ for the metal-poor GCs and $n(3D)=3.0$ for the metal-rich clusters.  The particle velocities were assigned randomly in a Gaussian distribution with a dispersion of 150 km s$^{-1}$ for all GCs, and their individual measurement uncertainties were randomly distributed uniformly from $5 - 80$ km s$^{-1}$.  All of these parameters mimic the NGC 5128 GC system.

For 1000 realizations, each GC system consisted of 550 objects, with $50\%$ red and $50\%$ blue subpopulations, which were projected from their real 3-dimensional positions into 2-dimensions.  We then searched for subgroups with our finding algorithm, as we have done for the observational data in Section~\ref{sec:obsdata}.  Our search criteria allowed objects to be group members only if they were within a projected length of 2\arcmin\ from each other, have a velocity difference of $<20$ km s$^{-1}$, exist outside of 10\arcmin\ from the galaxy's center, and have a velocity uncertainty less than 40 km s$^{-1}$ for each member.  Allowing the higher velocity uncertainty limit than imposed in most of our observational searches of 40 km s$^{-1}$ will only overestimate the number of false group members we find in the simulations.  Out of the 1000 trials, 9 of them detected one subgroup.  The subgroups consisted of 5-7 members.  This test therefore suggests that there is a $0.9\%$ chance of detecting a false group from our observationally mimiced dataset.

We further this work by generating a GC system that has 1300 GCs, which is the total estimated population in NGC 5128 \citep{harris10b}.  Our simulated system has 650 blue and 650 red clusters, projected into 2-dimensions.
In the observed GC system, there are 242 objects beyond our inner search radius of 10\arcmin\ so we randomly selected 242 objects beyond 10\arcmin\ from our simulated sample of 1300 GCs.  We repeated our group finding algorithm to this simulated system as described above.  This simulation represents the most realistic system we can make to match our observed dataset.  Out of our 1000 simulations, there were 3 trials that detected a subgroup of GCs, each of these groups consisted of 5 members, suggesting that we would find a fake subgroup of GCs with a $0.3\%$ chance.  Both this simulation, as well as our simulated group of 550 objects show that the probability of finding one fake subgroup of GCs in a system similar to our observed dataset is $< 1\%$. None of the trials detected more than one subgroup in the simulation.

\subsection{The Addition of Real Subgroups}
\label{sec:real}

It is important to also check how reliable our subgroup finding technique is at recovering a real group in our observed system.  This was tested by inserting a real subgroup into one of our simulations of 1300 members where no fake subgroup was found.  We generated 1000 real groups by randomly selecting their position in x, y, and z in the galaxy, as well as their velocity and velocity uncertainty.  Each real subgroup had between 4-6 members with a group velocity dispersion of within 20 km s$^{-1}$.  Their individual velocities were selected from a system with 150 km s$^{-1}$ dispersion.  The group was inserted between 10-50\arcmin\ from the galaxy's center which ensures the real group should be present in the projected system in our search regime.  The differences in the x and y and z positions for the group members was selected within 2\arcmin\ of each other.  

Out of the 1000 simulations that we performed, 892 of these simulations found all group members and no fake interlopers.  All real group members were recovered as well as the inclusion of fake members in 103 of the simulations.  In 2 of the simulations, some of the real group members were found with fake member interlopers, and in 3 of the simulations some of the group members were found with no fake interlopers detected.  We therefore consider our detection method to be very successful as in more than $>89\%$ of cases, we were able to find our exact simulation group system and $10.3\%$ of the time, we also found all group members however with interlopers.  When interlopers were present in the detected subgroups, $96.2\%$ of these occurrances had only 1 interloper, while 2 interlopers were detected in the remaining cases.  These results strongly suggest that at least some of our observationally detected subgroups are real and they cannot all simultaneously be false detections.  

\subsection{Testing for the Probability of Random Overlapping Groups}
\label{sec:overlap}

Lastly, we test whether our candidate subgroups might be smaller ones overlapped by projection purely by chance.  To the simulated GC system as a whole, we add random subgroups consisting of 4-6 members whose members are confined within a 2\arcmin\ region.  These subgroups were randomly placed in the galaxy between 10-50\arcmin\ before being deprojected into 2-dimensional space.  Two groups were considered to overlap if any one of their members were within a small designated distance from any member of another group.  The distance chosen did not significantly change our findings (considering differences between $0.1-2$\arcmin), so we selected 0.5\arcmin.  

We performed 7 simulations, each consisting of a different number of subgroups placed in the galaxy, which increased linearly from 2 subgroups up to 8 subgroups.  For each of these simulations, we examined, over 1000 trials, how many times these subgroups overlapped. The number of overlaps ranged from no overlapping subgroups up to 5 overlapping subgroups.  We did not find any simulation which had more than 5 overlapping groups. As would be expected, the higher the number of subgroups present, the higher the number of overlapping subgroups are found as a result. These probabilities of having random overlapping groups are listed in Table~\ref{tab:overlap}.  We find that the probability of having two subgroups randomly overlap is $\sim4\%$.   If all 8 subgroups in NGC 5128 (4 GCs and 4 PNe) are real groups, then the probability that 3 sets of subgroups randomly overlap is found to be $\sim7\%$.  Exluding Group 2, whose subgroups have very different average velocities, our results suggest there may be 2 sets of overlapping subgroups, Group 1 and Group3.  The probabilty of 2 sets of subgroups randomly overlaping with 8 subgroups present is $23\%$.  It is therefore not clear if these overlapping subgroups found in Groups 1 and 3 are real.

\section{Searching for Stellar Orbits}
\label{sec:orbits}

Work by \cite{peng02} shows a complex but faint shell structure threading the inner and mid-halo regions of NGC 5128.  These shells have been attributed to the phase-wrapping of stars from accreted satellites.  One particular blue elliptical arc was found to be associated with a young star clusters ($\simeq350$ Myr), but aside from this one example, no stellar streams have been associated with GCs or PNe.

We have begun a search for evidence of stellar streams in NGC 5128 by probing the objects within the galaxy that may trace the stream.  We use the available data for the GCs and the PNe, which may be stripped along with the stellar stream material during galaxy interactions.  If these objects are located, they can be used to identify the possible orbits of accreted satellites.
To do this, we use a friendless algorithm \citep{merrett03} and search for evidence of stellar streams using the GCs and PNe separately in the analysis below.

For each object in our sample with a measured radial velocity and a velocity uncertainty less than 50 km s$^{-1}$ (totalling 411 GCs and 780 PNe), we have found its $N$ nearest neighbours.  For those $N$ neighbours, we have calculated their mean velocity $v_{mean,N}$ and their velocity dispersion, $\sigma_N$.  We have classified the central object as friendless if its velocity is more than $n\times\sigma_N$ from $v_{mean,N}$.  
Varying the $N$ parameter does not seem to significantly alter the results.  We do find varying the $\sigma_N$ parameter from 3 to 4 changes the number of friendless objects by a few, while varying $\sigma_N$ from 2 to 3 does change the number of objects significantly.  

Taking a conservative approach and using $\sigma_N=4$, we find no friendless GCs when $N=30$, 1 when $N=20$ (GC0258), and 1 when $N=10$ (GC0578).  GC0258 will not be considered further as it was classified as a resolved GC in \cite{harris06} while a recent radial velocity measurement indicates this object may be a star \citep{woodley10b}, and it is because of this low radial velocity measurement that this object is identified.  GC0578 may indeed be friendless, but it is not possible to use one GC to indicate evidence for a stellar stream.  Using $N=10$ and $\sigma_N=3$ yields 7 friendless GCs (GC0216, GC0249, GC0325, GC0506, GC0577, GC0578, GC0580).  

We conduct the same analysis with the PNe dataset using $\sigma_N=4$ and find no friendless objects when $N=30$, 1 when $N=20$ (f05p02), and 6 when $N=10$ (4012, 4285, 5602, f05p02, f18p28, f42p10).  Using $N=10$ and $\sigma_N=3$ yields 20 friendless PNe (4012, 4128, 4285, 4417, 4511, 5203, 5601, 5602, 5617, 6104, f04p1, f05p02, f08p50=3002, f14p016, f18p28, f18p65=1208, f18p67=4023, f18p83, f42p10, mosNEpn4).  The locations of the inner $\sim 15$ of these are shown in Figure~\ref{fig:subgroups_all}.  

The friendless objects that are located at the largest distances from the galaxy's center are likely not indicators of streams.  Rather, there are very few neighbours as the incompleteness of planetary nebulae and GCs is quite high beyond 20\arcmin\ (see the radial distribution as a function of azimuth for the GCs in Figure 8 of \cite{woodley10b} and the distribution of the PN in Figure 2 of \cite{peng04a}).  This leaves only the inner 20\arcmin\ which can be explored for friendless objects, of which there are few (again, depending on the parameters chosen in the friendless algorithm).  We list these objects with the suggestion that (at least) some of them may trace stellar orbits, and continued searches for stellar streams or orbital modelling are necessary to go beyond this preliminary work. 

\section{Conclusions}
\label{sec:conclusions}

We have searched the GC and PN systems in the nearby giant elliptical galaxy, NGC 5128, for evidence of stellar subgroups and stellar streams from merging events.  Our results indicate there may be up to 4 potential subgroups of GCs and 4 potential subgroups of PNe. Two of the PNe subgroups overlap with two of the GC subgroups in position and velocity.  In order to improve the search for these subgroups in NGC 5128, we need higher precision radial velocity measurements to more concretely determine if these objects are grouped together.  

By generating GC systems that mimic that in NGC 5128, we are able to assign the probability of our results being a chance encounter due to the projection of these objects onto a 2-dimensional plane.  The probability of finding a fake subgroup within the GC population is $< 1\%$, while the probablity of having a set of subgroups overlap is $\sim 4\%$.  These results strongly suggest that at least some of these subgroups may indeed be real.  

We have also searched for evidence of stellar streams by locating GCs and planetary nebalae that differ from its neighbours in velocity space using a friendless search algorithm.  We present our findings as potential tracers, however further work in orbital modelling as well as searches for very faint surface brightness features would be necessary beyond this point to further define the stellar streams.

\acknowledgements K.A.W. thanks Dr. Harvey Richer and W.E.H. thanks NSERC for their financial
support.  K.A.W and W.E.H. thank Dr. Eric Peng for providing the image of NGC 5128 in Figure~\ref{fig:subgroups_all}.

%*************************REFERENCES*************************************
%***************************************************************************

\begin{deluxetable}{llllrrrr}
\tablecolumns{8}
\tabletypesize{\scriptsize}
\tablecaption{Possible Globular Cluster Subgroups in NGC 5128\label{tab:subgroups_gc}}
\tablewidth{0pt}
\tablehead{
\colhead{ID} & \colhead{Group ID}&\colhead{$R.A.$} & \colhead{$Decl.$}&
\colhead{[$Fe/H$]} & \colhead{$R_{gc}$}& \colhead{$v_r$} & \colhead{$\sigma_{v_r}$\tablenotemark{c}}  \\
\colhead{ } & \colhead{ }&\colhead{($J2000$)} & \colhead{($J2000$)}&
\colhead{ } & \colhead{(\arcmin)}& \colhead{(km s$^{-1}$)} & \colhead{(km s$^{-1}$)}  \\
}\startdata
GC0067                 &GC 1&13:24:51.49& -43:12:11.1& -1.17& 12.86& 624&  13\\
GC0083                 &GC 1&13:24:56.08& -43:10:16.4& -2.46& 10.79& 687&  31\\
GC0106                 &GC 1&13:25:01.83& -43:09:25.4& -1.42&  9.52& 688&  16\\
GC0114\tablenotemark{a}&GC 1&13:25:03.37& -43:11:39.6& -1.29& 11.41& 605&  46\\
GC0117                 &GC 1&13:25:04.48& -43:10:48.4& -0.32& 10.54& 626&  22\\
GC0122\tablenotemark{b}&GC 1&13:25:05.46& -43:14:02.6& -1.37& 13.51& 679&  13\\
GC0419                 &GC 1&13:24:55.31& -43:10:39.3& -0.35& 11.19& 656&  27\\
GC0422\tablenotemark{b}&GC 1&13:25:03.28& -43:08:14.4& -1.65&  8.37& 690&  31\\
GC0491                 &GC 1&13:24:54.98& -43:11:36.9& -0.22& 12.05& 670&  19\\
GC0492\tablenotemark{b}&GC 1&13:24:56.61& -43:12:23.6& -1.94& 12.59& 713&  28\\
GC0504\tablenotemark{b}&GC 1&13:25:08.94& -43:08:53.7& -1.96&  8.46& 678&  26\\
GC0354                 &GC 2&13:26:02.25& -43:08:55.6& -0.65& 10.03& 457&  38\\
GC0598                 &GC 2&13:26:09.52& -43:08:52.4& -0.41& 10.88& 482&  67\\
GC0379                 &GC 2&13:26:12.82& -43:09:09.2& -0.49& 11.50& 545&  60\\
GC0544                 &GC 2&13:26:17.27& -43:09:58.0& -0.05& 12.65& 473&  39\\
GC0393                 &GC 2&13:26:22.08& -43:09:10.7& -1.20& 12.79& 505&  78\\
GC0534                 &GC 3&13:25:55.35& -42:53:40.2& -2.30&  9.04& 347&  41\\
GC0590                 &GC 3&13:25:46.96& -42:52:34.0& -0.95&  9.28& 355&  66\\
GC0532                 &GC 3&13:25:54.79& -42:52:57.1& -0.33&  9.59& 399&  32\\
GC0341                 &GC 3&13:25:58.91& -42:53:18.9& -1.34&  9.70& 410&  20\\
GC0334                 &GC 3&13:25:56.59& -42:51:46.6& -0.73& 10.77& 403&  16\\
GC0518                 &GC 4&13:25:25.36& -42:49:27.5& -1.55& 11.70& 589&  63\\
GC0163                 &GC 4&13:25:15.79& -42:49:15.1& -1.44& 12.09& 534&  52\\
GC0516                 &GC 4&13:25:23.31& -42:48:47.7& -0.54& 12.38& 526&  11\\
GC0220                 &GC 4&13:25:30.72& -42:48:13.4& -1.88& 12.94& 484&  47\\
GC0511                 &GC 4&13:25:18.29& -42:47:52.4& -0.65& 13.39& 487&  39\\
\tablenotetext{a}{Included if velocity uncertainty cut-off increased from 40 km s$^{-1}$ to 50 km s$^{-1}$.}
\tablenotetext{b}{Included member in group if linking length increased from 2\arcmin\ to 3\arcmin.}
\enddata
\end{deluxetable}

\begin{deluxetable}{llllrrr}
\tablecolumns{7}
\tabletypesize{\scriptsize}
\tablecaption{Possible Planetary Nebulae Subgroups in NGC 5128\label{tab:subgroups_pne}}
\tablewidth{0pt}
\tablehead{
\colhead{ID} & \colhead{Group ID}&\colhead{$R.A.$} & \colhead{$Decl.$}&
 \colhead{$R_{gc}$}& \colhead{$v_r$} & \colhead{$\sigma_{v_r}$\tablenotemark{a}}  \\
\colhead{ } & \colhead{ }&\colhead{($J2000$)} & \colhead{($J2000$)}&
 \colhead{(\arcmin)}& \colhead{(km s$^{-1}$)} & \colhead{(km s$^{-1}$)\tablenotemark{a}}  \\
}\startdata
ctr513                 &PNe 1&13:26:14.26& -43:08:22.5 & 11.18& 480&  20\\
f08p10=5506            &PNe 1&13:26:06.19& -43:09:06.2 & 10.63& 462&  20\\
5508                   &PNe 1&13:26:13.01& -43:06:45.0 & 10.01& 439&  20\\
f07p06                 &PNe 1&13:26:11.33& -43:09:57.3 & 11.89& 421&  20\\
f07p08                 &PNe 1&13:26:06.87& -43:07:36.0 &  9.65& 476&  20\\
f08p9                  &PNe 1&13:25:56.04& -43:09:24.7 &  9.76& 439&  20\\
f17p34                 &PNe 2&13:26:11.29& -42:52:39.6 &11.65&  535&  20\\
f18p46                 &PNe 2&13:26:10.83& -42:54:01.5 &10.64&  527&  20\\
f18p51                 &PNe 2&13:26:07.42& -42:54:01.4 &10.18&  530&  20\\
f18p55=1217            &PNe 2&13:26:02.57& -42:53:07.1 &10.26&  516&  20\\
f18p74=1303            &PNe 2&13:26:05.35& -42:51:31.3 &11.84&  522&  20\\
5106                   &PNe 2&13:26:11.57& -42:55:37.7 & 9.75&  506&  20\\
5108                   &PNe 2&13:26:09.16& -42:55:20.1 & 9.56&  513&  20\\
f18p69=4031            &PNe 2&13:25:56.78& -42:53:50.4 & 9.05&  499&  20\\
f18p70                 &PNe 2&13:25:58.98& -42:53:39.0 & 9.44&  517&  20\\
f18p15                 &PNe 3&13:25:41.55& -42:45:41.5 &15.67&  519&  20\\
f18p16                 &PNe 3&13:25:38.18& -42:46:04.4 &15.20&  501&  20\\
f18p17                 &PNe 3&13:25:39.18& -42:45:24.2 &15.89&  534&  20\\
f18p18                 &PNe 3&13:25:34.32& -42:46:11.5 &15.01&  531&  20\\
f18p40                 &PNe 3&13:25:40.29& -42:43:44.0 &17.57&  553&  20\\
f18p26                 &PNe 3&13:25:26.79& -42:47:36.5 &13.54&  507&  20\\
ctr624                 &PNe 4&13:24:50.38& -42:53:22.6 &10.33  &557&  20\\
f19p06                 &PNe 4&13:24:45.76& -42:51:30.0 &12.31  &588&  20\\
f19p07                 &PNe 4&13:24:44.04& -42:52:09.9 &12.01  &538&  20\\
4608                   &PNe 4&13:24:49.87& -42:52:46.8 &10.85  &597&  20\\
ctr622                 &PNe 4&13:24:53.26& -42:53:26.8 & 9.94  &566&  20\\
\tablenotetext{a}{For all PNe, we have assumed a radial velocity uncertainty of 20 km s$^{-1}$.}
\enddata
\end{deluxetable}

\begin{deluxetable}{lrr}
\tablecolumns{3}
\tabletypesize{\scriptsize}
\tablecaption{Properties of the Subgroups\label{tab:properties}}
\tablewidth{0pt}
\tablehead{
\colhead{Group ID} & \colhead{$v_{mean}$}&\colhead{$\sigma$}  \\
\colhead{ } & \colhead{(km s$^{-1}$)}&\colhead{(km s$^{-1}$)}  \\
}\startdata
GC 1\tablenotemark{a}   &$651$&$25$\\
GC 2   &$492$&$26$\\
GC 3   &$382$&$19$\\
GC 4   &$524$&$34$\\
PNe 1  &$453$&$21$\\
PNe 2  &$518$&$11$\\ 
PNe 3  &$524$&$17$\\ 
PNe 4  &$569$&$21$\\ 
\tablenotetext{a}{These values have been calculated using the 7 probably group members of GC 1 using a linking length of 2\arcmin.} 
\enddata
\end{deluxetable}

\begin{deluxetable}{rrrrrrr}
\tablecolumns{7}
\tabletypesize{\scriptsize}
\tablecaption{Chance of Overlapping Subgroups\label{tab:overlap}}
\tablewidth{0pt}
\tablehead{
\colhead{N Subgroups}& \multicolumn{6}{c}{N Occurances out of 1000} \\
\colhead{ } & \colhead{0 overlaps}&\colhead{1 overlap} & \colhead{2 overlaps}&\colhead{3 overlaps}& 
\colhead{4 overlaps}&\colhead{5 overlaps}  \\
}\startdata
2  &958&  42&   0&  0&  0& 0\\
3  &869& 129&   2&  0&  0& 0\\
4  &783& 213&   4&  0&  0& 0\\
5  &631& 327&  40&  2&  0& 0\\
6  &439& 450&  99& 11&  1& 0\\ 
7  &358& 445& 160& 36&  1& 0\\ 
8  &249& 441& 229& 67& 13& 1\\  
\enddata
\end{deluxetable}


\begin{thebibliography}{}
\bibitem[Ashman \& Bird(1993)]{ashman93} Ashman, K.~M. \& Bird,
  C.~M. 1993, \aj, 106, 2281
\bibitem[Beasley et al.(2008)]{beasley08} Beasley, M. A., Bridges, T.,         
  Peng, E. W., Harris, W. E., Harris, G. L. H., Forbes, D. A., \&  
  Mackie, G. 2008, \mnras, 386, 1443            
\bibitem[Bellazzini et al.(2003)]{bellazzini03} Bellazzini, M., Ferraro, F.~R., \& Ibata, R.~A. 2003, \aj, 125, 188
\bibitem[Bullock \& Johnston(2005)]{bullock05} Bullock, J.~S. \& Johnston, K.~V. 2005, \apj, 635, 931
\bibitem[Buzzoni et al.(2006)]{buzzoni06} Buzzoni, A., Arnaboldi, M., \& Corradi, R. L. M. 2006, MNRAS, 368, 877
\bibitem[Ciardullo et al.(2005)]{ciardullo05} Ciardullo, R., Sigurdsson, S., Feldmeier, J. J., \& Jacoby, G. H. 2005, \apj, 629, 499
\bibitem[C\^ot\'e et al.(1998)]{cote98} C\^ot\'e, P., Marzke, R. O., \& West, M. J. 1998, \apj, 501, 554 
\bibitem[C\^ot\'e et al.(2002)]{cote02} C\^ot\'e, P., West, M. J., \& Marzke, R. O. 2002, \apj, 567, 853
\bibitem[Dufour et al.(1979)]{dufour79} Dufour, R. J., van den Bergh,
  S., Harvel, C. A., Martins, D. H., Schiffer, F. H., Talbot, R. J.,
  Talent, D. L., \& Wells, D. C. 1979, \aj, 84, 284 
\bibitem[Ellis et al.(2009)]{ellis09} Ellis, S. et al. 2009, The Many Faces of Cen A, Conference
\bibitem[Forbes et al.(2003)]{forbes03} Forbes, D. A., Beasley, M.~A., Bekki, K., Brodie, J.~P., \& Strader, J. 2003, Science, 301, 1217
\bibitem[Forte et al.(1982)]{forte82} Forte, J.~C., Mart{\'i}nez, R.~E., \& Muzzio, J.~C. 1982, \aj, 87, 1465
\bibitem[Geha et al.(2010)]{geha10} Geha, M., van der Marel, R. P., Guhathahurta, P., Gilbert, K. M., Kalirai, J., \& Kirby, E. N. 2010, \apj, 711, 361
\bibitem[Graham(1979)]{graham79} Graham, J. A. 1979, \apj, 232, 60
\bibitem[Harris \& van den Bergh(1981)]{harris81} Harris, W. E., \& van den Bergh, S. 1981, \aj, 86, 1627
\bibitem[Harris et al.(1992)]{harris92} Harris, G.~L.~H., Geisler, D.,
  Harris, H.~C., \& Hesser, J.~E. \aj, 104, 613
\bibitem[Harris \& Harris(2002)]{harris02} Harris, W.~E. \& Harris, G.~L.~H. 2002, \aj, 123, 3108
\bibitem[Harris et al.(2004b)]{harris04b} Harris, G. L. H., Harris,
  W. E., \& Geisler, D. 2004b, \aj, 128, 723
\bibitem[Harris et al.(2006)]{harris06} Harris, W. E., Harris, G. L. H., Barmby, P., McLaughlin, D. E.. \& Forbes, D. A., 2006, \aj, 132, 2187
\bibitem[Harris et al.(2010a)]{harris10} Harris, G. L. H., Rejkuba,
M. \& Harris, W. E. 2010a, arXiv0911.3180
\bibitem[Harris(2010b)]{harris10b} Harris, G. L. H. 2010b, arXiv1004.4907  
\bibitem[Hesser et al.(1984)]{hesser84} Hesser, J. E., Harris, H. C.,
  van den Bergh, S., \& Harris, G. L. H. 1984, \apj, 276, 491
\bibitem[Hesser et al.(1986)]{hesser86} Hesser, J. E., Harris, H. C.,
  \& Harris, G. L. H. \apj, 303, L51
\bibitem[Hilker et al.(1999)]{hilker99} Hilker, M., Infante, L., \& Richtler, T. 1999, \aaps, 138, 55
\bibitem[Hui et al.(1995)]{hui95} Hui, X., Ford, H.~C., Freeman, K.~C., \&
  Dopita, M.~A. 1995, \apj, 449, 592
\bibitem[Ibata et al.(1994)]{ibata94} Ibata, R.~A., Gilmore, G., \& Irwin, M.~J. 1994, \nat, 370, 194
\bibitem[Ibata et al.(1997)]{ibata97} Ibata, R.~A., Wyse, R.~F.~G., Gilmore, G., Irwin, M.~J., \& Suntzeff, N.~B. 1997, \aj, 113, 634
\bibitem[Ibata et al.(2001a)]{ibata01a} Ibata, R.~A., Lewis, G. F., Irwin, M., Totten, E., \& Quinn, T. 2001a, \apj, 551, 294 
\bibitem[Ibata et al.(2001b)]{ibata01b} Ibata, R.~A., Irwin, M., Lewis, G., Ferguson, A. M. N., \& Tanvir, N. 2001b, \nat, 412, 49
\bibitem[Jacoby(1980)]{jacoby80} Jacoby  G., 1980, \apjs, 42, 1 
\bibitem[Lynden-Bell \& Lynden-Bell(1995)]{lyndenbell95} Lynden-Bell, D. \& Lynden-Bell, R.~M. 1995, \mnras, 275, 429
\bibitem[Mackey et al.(2010)]{mackey10} Mackey, D. et al. 2010 (in prep)
\bibitem[Majewski et al.(2003)]{majewski03} Majewski, S.~R., Skrutskie, M.~F., Weinberg, M.~D., \& Ostheimer, J.~C. 2003, \apj, 599, 1082
\bibitem[Malin et al.(1983)]{malin83} Malin, D.~F., Quinn, P.~J., \& Graham, J.~A. 1983, \apjl, 272, 5
\bibitem[Malin \& Hadley(1997)]{malin97} Malin, D. \& Hadley, B. 1997, PASA, 14, 52
\bibitem[Mart{\'{\i}}nez-Delgado et al.(2004)]{martinez04} Mart{\'{\i}}nez-Delgado, D., G{\'{o}}mez-Flechoso, M. A., Aparicio, A., \& Carrera, R. 2004, \apj, 601, 242
\bibitem[Mart{\'{\i}}nez-Delgado et al.(2008)]{martinez08} Mart{\'{\i}}nez-Delgado, D., Pe{\~n}arrubia. J., Gabany, R.~J., Trujillo, I., Majewski, S.~R., \& Pohlen, M., 2008, \apj, 689, 184
\bibitem[Mateo(1998)]{mateo98} Mateo, M.~L. 1998, \araa, 36, 435
\bibitem[McConnachie et al.(2009)]{mcconnachie09} McConnachie, A. W. et al. 2009, \nat, 461, 66
\bibitem[Merrett et al.(2003)]{merrett03} Merrett, H.~R., Kuijken, K., Merrifield, M.~R., Romanowsky, A.~J., Douglas, N.~G., Napolitano, N.~R., Arnaboldi, M., Capaccioli, M., Freeman, K.~C., Gerhard, O., Evans, N.~W., Wilkinson, M.~I., Halliday, C., Bridges, T.~J., \& Carter, D. 2003, \mnras, 346, L62 
\bibitem[Mouhcine et al.(2010b)]{mouhcine10b} Mouhcine, M., Harris, W.~E., Ibata, R., \& Rejkuba, M. 2010b, \mnras, 404, 1157
\bibitem[Mouhcine et al.(2010a)]{mouhcine10a} Mouhcine, M., Ibata, R., \& Rejkuba, M. 2010a, \apjl, 714, 12
\bibitem[Muzzio(1987)]{muzzio87} Muzzio, J.~C. 1987, \pasp, 99, 245
\bibitem[Peng et al.(2002)]{peng02} Peng, E.~W., Ford, H.~C.,
  Freeman, K.~C. \& White, R.~L. 2002, \aj, 124, 3144
\bibitem[Peng et al.(2004a)]{peng04a} Peng, E.~W., Ford, H.~C., \&
  Freeman, K.~C. 2004a, \apj, 602, 685
\bibitem[Peng et al.(2004b)]{peng04b} Peng, E. W., Ford, H. C., \&
  Freeman, K. C. 2004b, \apj, 602, 705
\bibitem[Peng et al.(2008)]{peng08} Peng, E. W., Jord{\'a}n, A., C{\^o}t{\'e}, P., Takamiya, M., West, M. J., Blakeslee, J. P., Chen, C.-W., Ferrarese, L., Mei, S., Tonry, J. L., \& West, A. A. 2008, \apj, 681, 197
\bibitem[Perrett et al.(2003)]{perrett03} Perrett, K. M., Stiff, D.~A., Hanes, D.~A., \& Bridges, T.~J. 2003, \apj, 589, 790
\bibitem[Pipino et al.(2007)]{pipino07} Pipino, A., Puzia, T.~H., \& Matteucci, F. 2007, \apj, 665, 295
\bibitem[Reed et al.(1994)]{reed94} Reed, L.~G., Harris, G.~L.~H., \& Harris, W.~E. 1994, \aj, 107, 555
\bibitem[Rejkuba et al.(2001)]{rejkuba01} Rejkuba, M., Minniti, D., Silva, D.~R., \& Bedding, T.~R. 2001, \aap, 379, 781
\bibitem[Rejkuba et al.(2002)]{rejkuba02} Rejkuba, M., Minniti, D., Courbin, F., \& Silva, D.~R. 2002, \apj, 564, 688
\bibitem[Rejkuba et al.(2007)]{rejkuba07} Rejkuba, M., Dubath, P.,
  Minniti, D., \& Meylan, G. 2007, \aap, 469, 147
\bibitem[Schiminovich et al.(1994)]{schiminovich94} Schiminovich, D.,
  van Gorkum, J. H., van der Hulst, J. M., \& Kasow, S. 1994, \apj,
  423, L101
\bibitem[Shang et al.(1998)]{shang98} Shang, Z., et al. 1998, \apj, 504, L23
\bibitem[Sharina et al.(2010)]{sharina10} Sharina, M. E., Chandar, R., Puzia, T.~H., Goudfrooij, P., \& Davoust, E. 2010, \mnras, tmp, 536 (NEED VOLUME)
\bibitem[van den Bergh et al.(1981)]{vandenbergh81} van den Bergh, S.,
  Hesser, J.~E., \& Harris, G.~L.~H. 1981, \aj, 86, 24
\bibitem[Woodley et al.(2005)]{woodley05} Woodley, K. A., Harris,
  W. E. \& Harris, G. L. H. 2005, \aj, 129, 2654 
\bibitem[Woodley et al.(2007)]{woodley07} Woodley, K.~A., Harris, W.~E., Beasley, M.~A., Peng, E.~W., Bridges, T.~J., Forbes, D.~A., \& Harris, G.~L.~H. 2007, \aj, 134, 494
\bibitem[Woodley et al.(2010a)]{woodley10a} Woodley, K.~A., Harris,
  W.~E., Puzia, T.~H., G{\'o}mez, M., Harris, G.~L.~H., \& Geisler,
  D. 2010a, \apj, 708, 1335
\bibitem[Woodley et al.(2010b)]{woodley10b} Woodley, K.~A., G{\'o}mez, M., Harris,  W.~E., Geisler, D., \& Harris, G.~L.~H.,  2010b, \aj, 139, 1871
\bibitem[Yoon \& Lee(2002)]{yoon02} Yoon, S.-J. \& Lee, Y.-W. 2002, Science, 297, 578

\end{thebibliography}
\end{document}